\documentstyle[preprint,aps,prbbib]{revtex}
\draft
\begin{document}
\title{Electric fields and valence band offsets at
strained [111] heterojunctions}

\author{S.Picozzi and A. Continenza}
\address{INFM - Istituto Nazionale di Fisica della Materia\\
Dipartimento di Fisica \\
Universit\`a degli Studi di L'Aquila, 67010 Coppito (L'Aquila), Italy \\ and}
\author{A.J.Freeman}
\address{Department of Physics and Astronomy and Material Research Center\\
Northwestern University, Evanston, IL 60208 (U.S.A.)\\}

\maketitle
\narrowtext

\begin{abstract}
[111] ordered
common atom strained layer superlattices
(in particular the common anion
GaSb/InSb system and the common cation InAs/InSb
system) are investigated using the {\em ab initio} full potential
linearized augmented plane wave (FLAPW) method.
We have focused our
attention on the potential line-up at the two sides of the homopolar isovalent
heterojunctions considered, and in particular on its dependence on the strain
conditions and on the strain induced electric fields. We propose a
procedure to locate the interface plane where the band alignment could be
evaluated; furthermore, we suggest that the polarization
charges, due to piezoelectric effects, are approximately confined to a narrow
region close to the interface and
do not affect the potential discontinuity.
We find  that the interface contribution to the
valence band offset is
substantially unaffected by strain conditions, whereas
 the total
band line-up is highly tunable, as a function of the  strain conditions.
Finally, we compare our
results with those obtained for [001] heterojunctions,
\end{abstract}

\section{Introduction}

As is well known, the opportunity of tuning the
potential line-up as a  function of the strain has a relevant
 technological importance and thus has been the subject of many recent
experimental and theoretical works.
\cite{maria,gorczyca,swaz,ohler,silvia2,vandewalle} However,
most of them focused on
CuAu-like
strained interfaces ({\em i.e.} oriented along the [001]  direction).
In this work we have performed
{\em ab initio} full-potential augmented plane waves (FLAPW) \cite{FLAPW}
calculations (within density functional theory \cite{dft}) for
GaSb/InSb and InAs/InSb  systems, in order to
examine the valence band offset (VBO)
at strained [111] heterojuctions.
To our knowledge,
only recently have CuPt-like ({\em i.e.} [111] ordered) systems
begun to be investigated
\cite{froyen,harshman} and
the way the charge readjustment occurs
at the 
interface is still far from being clear, mostly
 due to the non-trivial
electrostatics involved.


Although both the [001] and [111] directions are polar (the
[110] is
non-polar), there are important differences between interfaces
oriented along these two crystallographic axes.
First of all, the particular geometry for the [111] superlattice leads to two
inequivalent interfaces: \cite{bbr} consider
the structure for the (GaSb)$_3$/(InSb)$_3$ superlattice shown in Fig.
\ref{fig1}
(note that the common cation case has the same structure, except for the atomic
species involved).
According to convention, we have placed the interface planes
$\pi_c$ ($\pi_s$) at the center (sides)
of the interface InSb (GaSb) axial bond (as shown in Fig. \ref{fig1}).
Observe that the interface Sb is coordinated with three Ga atoms and one In atom
in one interface ($\pi_{c}$ in Fig. \ref{fig1}),
and with three In atoms and one Ga atom in the other ($\pi_{s}$ in Fig.
\ref{fig1}).
Thus,  we could have, in principle,
two different valence band offsets and charge
accumulations ($\sigma$) at the interface planes $\pi_{c}$ and $\pi_s$,
namely $\sigma_{c}=-\sigma_{s}$, because of charge neutrality
requirements. \cite{bbr}
Incidentally, we note that
this is not the case for the [001] ordered systems, where the
bond directions and the
atomic coordinations result in exactly equivalent interfaces
at the center and at the sides of the unit cell.

Furthermore, since the semiconductor constituents
(with zincblende structure) are
piezoelectric materials, \cite{martin,degiro} a uniaxial strain
along the [111] direction
leads to a strain induced piezoelectric polarization ${\bf P_s}$.
\cite{harshman,nye,dalcorso} Following the notation of Ref.
\onlinecite{dalcorso},
we obtain:
\begin{equation}
|{\bf P}_s| \:= P_s = -\sqrt3 \: \gamma_{14} \:\epsilon_{4}
\label{ppiezo}
\end{equation}
 where $\gamma_{14}$ is the piezoelectric constant and
$\epsilon_{4}$ is the off-diagonal strain tensor element.
As a function of the in-plane strain ($\epsilon_{\perp} =
\frac{a_{sub}-a_{eq}}{a_{eq}}$,
with $a_{sub}$ and $a_{eq}$ denoting the substrate and equilibrium bulk lattice
constants, respectively) and within the linear regime, \cite{dalcorso} the
strain tensor component can be expressed as:
\begin{equation}
\epsilon_{4} \:=-2 \frac{c_{11}+2\:c_{12}}
{c_{11}+2\:c_{12}\:+ 4\:c_{44}} \epsilon_{\perp}
\end{equation}

The piezoelectric effect
can generate electric
fields so that: \cite{mailhotrmp}
\begin{equation}
{\bf P}\:={\bf P_s}+\varepsilon_0 \chi \:{\bf E}
\label{ps}
\end{equation}
where ${\bf P}$ is the dielectric polarization and
$\chi$ is the
dielectric susceptibility.
Note that for strained layer superlattices (SLs) with [001] growth axis,
the strain tensor is diagonal, so that
${\bf P_s}$ vanishes. 
 \cite{mailhotrmp}

In this paper,
we discuss results obtained for common--cation and common--anion
strained layer superlattices grown along the [111] direction and
discuss the effects of the strain induced polarization on the properties
of the superlattices.
In Sect. II we give some calculational details; in Sect. III and IV
we briefly review the electrostatics involved; in Sect. V we discuss
how charge redistributes at the interface and how the VBO
 can be evaluated. Finally, in Sect. VI we summarize our
conclusions.

\section{Structural and technical details}

We have studied the  SL systems (GaSb)$_3$/(InSb)$_3$
and  (InAs)$_3$/(InSb)$_3$ with unit cells containing 12 atoms
(as shown in Fig. \ref{fig1}).  For this
we have considered an AC/BC type heterojunction
(the case of the common anion AB/CB type heterojunction is exactly equivalent)
grown on different substrates:
($i$) on a bulk AC semiconductor ({\em i.e.} $a_{sub} = a_{AC}$, where
$a_{AC}$ denotes the AC semiconductor equilibrium lattice
constant); ($ii$)
 on a bulk BC semiconductor ({\em i.e.}
$a_{sub} = a_{BC}$); ($iii$) on a substrate having an average
lattice parameter ({\em i.e.} $a_{sub} = \frac{1}{2} (a_{AC} + a_{BC}$)).
The structural parameters for the different systems considered were determined
according to the macroscopic theory of elasticity (MTE) and are equal to those
for the ultrathin 1x1 [111] ordered SLs of Ref. \onlinecite{silvia1}.
With this choice, we are not considering any internal strain \cite{martin1}
which would produce different bond lengths
\cite{vdw2} for bonds parallel to the [111] and the other directions.
\cite{nota} 
Of course, we expect that variations in the bond lengths may
greatly affect the magnitude of the electric fields; very 
large deviations may even cause a sing change.
However, 
all this will not change the basic electrostatics involved and,
as we will show, the physical quantities we are interested in, such
as the VBO.

We performed all-electron full-potential linearized augmented plane
wave (FLAPW) calculations within the local density approximation
(LDA) \cite{dft}
to density functional theory with the exchange and correlation potentials
as parametrized by
Hedin and Lundqvist. \cite{HL}
Technical details (such as wave function cutoff, special k-points mesh etc.)
are equal to those  used in a previous study and  reported elsewhere.
\cite{silvia2,silvia1} In order to check
the convergence of the valence band offset
as a function of the cell dimension, we also
performed calculations for
(GaSb)$_4$/(InSb)$_4$
and  (InAs)$_4$/(InSb)$_4$ SLs grown on the same substrates.


\section{Strain induced electric fields}

Since they are important quantities for determining valence band
offsets, we report in Fig. \ref{fig2} 
(Fig. \ref{fig3})
the common atom Sb (In) core level binding energies
$E^b_{CL}$ for the 
GaSb/InSb (InAs/InSb) [111] ordered SLs grown on an average
substrate.
In the upper part of Fig. \ref{fig3} (a), we also report the core level
binding energies for 3x3 [001] ordered InAs/InSb grown on an average
substrate.

First of all, we point out that while in the 4x4 structures
 the $E^b_{CL}$
are perfectly aligned on a straight line in the bulk region, the same
behaviour is not so obvious to extrapolate
in the 3x3 structures, due to the smaller
bulk region. However,
in what follows, we will assume the linear behaviour also
for these structures and verify the validity of this assumption afterwards,
when comparing the VBOs obtained.

From Figs. \ref{fig2} and \ref{fig3}, we observe that the
structures considered show electric fields 
with opposite signs at the two sides
of the interface.
The slopes of the linear trends of Figs. \ref{fig2} and \ref{fig3}
give a rough estimate of the absolute values of the electric fields
($|{\bf E}|\:=\:E=\frac{\Delta\:E^b_{CL}}{\Delta\:z}$).
Our calculations give about 3x10$^7$ V/m (12x10$^7$ V/m) in both the
GaSb (InAs) and InSb sides of the common anion (cation) heterojunction,
with an error bar $\simeq$
2x10$^7$ V/m.
Note that this large error bar
is due to the very small variations
(in particular for the common anion case)
of $E^b_{CL}$ in the bulk region,
which are of the same order
of magnitude of our numerical uncertainty
upon the core level binding energies themselves.

Furthermore, we observe that the plot
of the core level binding energies
for [001] strained layer SLs
leads to a constant trend (implying a constant value of $E^b_{CL}$),
as the
$z$ coordinate 
(perpendicular to the interface) is varied within each
semiconductor bulk region; this confirms the absence of electric fields in
the [001] ordered structures.

\section{Piezoelectricity and polarization charges}

As already pointed out, \cite{mailhotrmp} [111] ordered strained
SLs show electric fields, due to piezoelectricity. Thus, elementary
electrostatics of dielectric media leads to the following relation between
the macroscopic quantities
in
each semiconductor constituent $(i)$:
\begin{equation}
{\bf D}^{(i)}\:=\varepsilon_0\:{\bf E}^{(i)}+{\bf P}^{(i)}=
\varepsilon_0\:{\bf E}^{(i)}+
\varepsilon_0 \chi^{(i)} {\bf E}^{(i)}+ {\bf P}_s^{(i)}
\label{deq}
\end{equation}
Due to symmetry properties, the transverse ({\em i.e.} parallel to the
interface) component of the piezoelectric polarization vector
vanishes; thus, in the following, we will only consider
the longitudinal ({\em i.e.} perpendicular to the  interface) component
of the electrostatic quantities involved.

The net charge accumulation at an ideal abrupt interface is related to the
macroscopic polarizations in the two semiconductor constituents, through
the following relation: \cite{mailhotrmp}

\begin{equation}
\sigma= - ({\bf P}^{(2)} -{\bf P}^{(1)}) \cdot \hat{{\bf n}}
\label{sigma}
\end{equation}
Since the perpendicular component $D_n$ is continuous across the interface,
Eq. (\ref{deq}) gives:
\begin{equation}
\varepsilon_0 {\bf E}^{(1)} + {\bf P}^{(1)} =
\varepsilon_0 {\bf E}^{(2)} + {\bf P}^{(2)}
\label{ugua}
\end{equation}
so that Eq.(\ref{sigma}) can be rewritten as:
\begin{equation}
\sigma= + \varepsilon_0 ({\bf E}^{(2)} -{\bf E}^{(1)}) \cdot \hat{{\bf n}}
\label{eqsigma1}
\end{equation}
In addition, from Eq.(\ref{sigma}) and using Eq.(\ref{ps}), we obtain an
equivalent
expression for $\sigma$:
\begin{equation}
\sigma = [ + \varepsilon_0 \:(\chi^{(2)} {\bf E}^{(2)} 
- \chi^{(1)} {\bf E}^{(1)}) + {\bf P_s}^{(2)}
 -{\bf P}_s^{(1)}\:]
\cdot \hat{{\bf n}}
\label{eqsigma2}
\end{equation}
Note that, in the case of growth on a substrate
equal to one of the constituent materials,
only one side of the interface is strained and gives
rise to a non-vanishing strain polarization
${\bf P_s}$. 
On the other hand, in the case of growth on an average
substrate, ${\bf P_s}$ is non-zero and will have
opposite  signs in the two different materials. In
fact, the piezoelectric constants \cite{degiro} have the same sign for all the
materials involved, while the two constituents are in biaxial tension and
compression, respectively, leading to in-plane strains
$\epsilon_{\perp}^{(1)}$
and $\epsilon_{\perp}^{(2)}$, with opposite signs.

In order to test the validity of the above description, we have
performed the double macroscopic average ($\overline{\overline{\rho}}^T$)
 \cite{bbr,ale90} of the self consistent
total (electrons + ions)
charge density ($\rho^T = \rho^{el} + \rho^{ion}$)
for 
some of the cases considered.
We have estimated the net interface charge accumulation from the following
relation:
\begin{equation}
\sigma_{SCF}=
\int_{b_1}^{b_2} \overline{\overline{\rho}}^T(z) dz
\label{mediamacro}
\end{equation}
where 
the integration is performed  between the two bulk regions  (referring to Fig.
\ref{fig1} for the 3x3 case, between
$b_1$ and $b_2$).

We must point out that the use of the symbol $\sigma$ for the charge
accumulation does not  necessarily imply that this charge must have a
surface-like distribution. In fact, the electrostatics used so far, holds for
any volume distributed charge density. However, as will be discussed later, the
charge distribution at the interface can be reasonably approximated as a
planar charge.

Using Eq. (\ref{mediamacro}) and Eq. (\ref{eqsigma1}), we can check the
consistency of our results. In the (GaSb)$_4$/(InSb)$_4$ SL grown
on an average substrate, for example,
we obtain $\sigma_{SCF} \: \approx$   2 x 10$^{-4}$ C/m$^2$
from Eq. (\ref{mediamacro}). Using Eq. (\ref{eqsigma1}) and the electric fields
as evaluated by the core level binding energies
($E^{(1)} \approx$ 2.6 x 10$^7$ V/m in the GaSb side of the heterojunction and
$E^{(2)} \approx$ 3.3 x 10$^7$ V/m  in the InSb side of the heterojunction),
we find
$\sigma \approx$ 5  x 10$^{-4}$ C/m$^2$.
The  agreement
between this value and $\sigma_{SCF}$ is within our numerical accuracy,
which is estimated to be 1.8 $\cdot$ 10$^{-4}$ C/m$^2$
(considering the error bar on the electric fields) and 
shows the consistency of our calculations.

However, if one tries to estimate  ${\bf P}_s^{(1)}$ and ${\bf P}_s^{(2)}$
from
Eq. (\ref{ppiezo})  
and then substitutes the results
in Eq. (\ref{eqsigma2}),
a value for $\sigma \: \approx$  49 x 10$^{-4}$ C/m$^2$ 
($\sigma \: \approx$  18 x 10$^{-4}$ C/m$^2$) 
is found, using the available calculated \cite{degiro} (experimental
\cite{harrison,martin})
elastic, dielectric  and
piezoelectric
constants  for each material. These values
 seem to be at variance with the integrated value $\sigma_{SCF}$. The
disagreement can be attributed to the
use of bulk lattice constants for a region whose thickness is as small as a few
monolayers and to the fact that the theoretical
and experimental constants
\cite{degiro,martin} used in Eq. (\ref{eqsigma2})
are not consistent with the present calculation, 
since the MTE structures
considered do not properly take into account internal strain
effects.
Let us remark that the inconsistency between these values of $\sigma$ and
the one obtained {\em ab-initio}
will not affect the evaluation of the VBO, 
as will be shown (see discussion below). 

Furthermore, from the steeper slopes in Fig.
\ref{fig3} with respect to those in
Fig. \ref{fig2}, we observe that the magnitudes of
the electric fields in the common cation systems are stronger than
those in the common  anion ones.
Since Eq. (\ref{eqsigma1}) leads to a net charge accumulation
proportional to the discontinuity of the electric fields,
it is reasonable to expect a
systematically  smaller charge,
$\sigma$, in the GaSb/InSb than in the
InAs/InSb systems, as confirmed by our self-consistent results
obtained from Eq. (\ref{mediamacro}).

\section{Valence Band Offset}

We recall that the potential line-up can be estimated following the procedure
used in XPS experiments and widely adopted in all-electron calculations,
\cite{vboaz,mmf} which takes core level
binding energies as reference levels.
According to this method, the VBO is obtained as follows:
\begin{equation}
\Delta\:E_v\:=\Delta\:b\:+\Delta\:E_b
\label{VBO}
\end{equation}
Here $\Delta\:b$ is an ``interface" term (which denotes the
core level energy difference between  atoms in the two bulk regions,
{\em i.e.} those denoted by ``b$_1$" and ``b$_2$" in Fig.
\ref{fig1})
and  $\Delta\:E_b$
is the so called
``bulk" term and indicates the binding energy difference - with respect
to the topmost valence level - between the same core levels in each
semiconductor strained as in the heterojunction.
This last term takes into account the effects of strain on the electronic
band structure of each bulk material.

We must now note that the evaluation of the $\Delta\:b$ term is usually done
assuming that the bulk regions of the SL are thick enough so that the atoms
taken as reference can be, by all means, considered as ``bulk"
atoms.  
This procedure implicitly requires that the $\Delta\:b$ term must be
evaluated in the limit of infinite distances from the interface, where
the reference atoms are no longer influenced by the charge rearrangement at the
interface.

It is now obvious that this same procedure cannot be used for the   [111]
oriented heterojunctions. The presence of electric fields, in fact, makes
the $\Delta\:b$ term vary with the distance from the interface, so that
its evaluation in the limit of ``bulk region" ({\em i.e.} infinitely
far away from the interface) would result in an ill-defined $\Delta\:b$.
Let us observe, by the way, that  the same conceptual problem is present
for procedures that use the electrostatic potential, rather than core levels,
as reference energies for the evaluation of the VBO. \cite{bbr}
This problem has been usually overcome \cite{froyen,fiore} by extrapolating
graphically the linear behaviour of the potential (or the core level
binding energy) in the two bulk regions to an ``interface plane", $\pi$,
arbitrarily taken half-way along the interface bond, and evaluating
the difference between their intercepts.

In the following, we illustrate a more rigorous procedure which will
also be able to give a reasonable estimate of the related error bar.

\subsection{Interface charge and evaluation of the VBO}

Let us first discuss how the charge is modified at the interface.
We can split up
this charge distribution
as the sum of
two different effects: ({\em i})
a charge accumulation $\sigma$, due to the piezoelectric effect combined
with the
boundary conditions, which is basically a monopole-like distribution
and ({\em ii})  a charge
depletion and accumulation at the two sides of the heterojunction,
which is the usual dipole charge of the interface
and is due to the different chemical nature and electronegativity of the
constituents.
The charge profile can thus be schematically represented
as a symmetric
monopole centered on some ideal interface plane $\pi$ plus a dipole
which is anti-symmetric with respect to $\pi$.

According to this scheme
and following basic electrostatic arguments, we can now say that
the monopole charge $\sigma$
leads to a continuous potential across the interface plane.
The limits for $z \rightarrow \infty$ and $z \rightarrow - \infty$ of this
potential must be the same, since the monopole contribution is symmetric with
respect to the plane $\pi$. On the other hand,
the dipole-like
charge leads, in these same limits,  to a potential
discontinuity. Thus, we expect that
the valence band offset ({\em i.e.} the potential discontinuity at the
interface) is only determined by the usual dipole term and
is not affected by the presence of the
strain induced electric fields.

Note that the real total
charge density macroscopic average profile ($\overline{\overline{\rho}}^T$)
usually has a very complex shape, often changing its sign.
Therefore, it is not possible to clearly establish
where $\pi$ 
really is.
However, as a rough approximation, we can calculate
the  center of gravity of the positive ($\overline{\overline{\rho}}^T_+(z)$)
 and negative ($\overline{\overline{\rho}}^T_-(z)$)
charges:
\begin{equation}
\overline{z}_+ =\frac{\int z \overline{\overline{\rho}}^T_+(z) dz}
{\int \overline{\overline{\rho}}^T_+(z) dz}\\
\end{equation}
\begin{equation}
\overline{z}_- =\frac{\int z \overline{\overline{\rho}}^T_-(z) dz}
{\int \overline{\overline{\rho}}^T_-(z) dz}\\
\end{equation}
where $\overline{\overline{\rho}}^T(z) = \overline{\overline{\rho}}^T_+ (z)
+\overline{\overline{\rho}}^T_-(z)$ and
the integrals are performed between the two bulk regions $b_1$ and $b_2$
as in Fig. \ref{fig1}.
(\ref{mediamacro}).  We may now  assume that
the interface region is confined between the two extremes
$\overline{z}_+$ and $\overline{z}_-$.





We are now able to calculate the $\Delta\:b$ term appearing in Eq.
(\ref{VBO}) and therefore the VBO.
A reasonable estimate for the $\Delta\:b$ term could then be obtained
from the intercepts at $z_{+}$ and $z_-$ of the two straight
lines which interpolate the core level energies.
The difference of these two values can
be taken as the error bar. We find, for the systems considered, that the
values of  $\overline{z}_+$ and $\overline{z}_-$ are, as expected, not
symmetric with respect to the half-bond position (which is, in any case,
within the ($\overline{z}_+, \overline{z}_-$) interval), due to two different
effects: ({\em i}) the strain, which modifies the bond lengths at the interface
and ({\em ii)} the distribution and sign of the monopole term $\sigma$.
In
all the cases considered, $\overline{z}_+$ and $\overline{z}_-$ differ by
as much
as 2 a.u., which is to be compared with bond lengths along the $z$
direction that
are usually of the order of 5 a.u.. This estimate of the charge centers of
gravity confines, with some precision, the ``interface" to a quite localized
region, which is even smaller than a bond length. Within this error, we may
also say that the interface charge has a surface-like distribution. The
numerical uncertainty on the $\Delta\:b$ term can thus be related to the
extrapolation procedure on the two extremes of the interface region,
$\overline{z}_+$ and $\overline{z}_-$, leading to a numerical uncertainty of
about 0.03 eV.

We may now recall, as already pointed out, that in a [111] ordered SL,
we have two inequivalent interfaces with, in principle, different
$\Delta\:b$'s.
For each of the structures examined, 
we find that
the
$\Delta\:b$ values obtained following this procedure at the two
inequivalent  interfaces differ by as  much as 0.02 eV,
so that we can consider, as a first approximation,
$\Delta\:b_{c}=\Delta\:b_{s}$.

Furthermore, we
also obtained equal $\Delta\:b$ values  for the 3x3 and
4x4 SLs (within 0.02 eV), thus confirming
the validity of the extrapolation procedure; this also suggests
that bulk conditions between the two
interfaces are recovered even in the smaller unit cell and
that good convergence for
the $\Delta\:b$ term is reached, as a function of the cell dimensions.
Note that the value (0.17 eV) obtained in Ref. \onlinecite{silvia2} for
InAs/Insb [111] grown on an average substrate
is in good agreement with the one derived from
Fig. \ref{fig3} (a) (0.19 eV). The
small discrepancy (about 0.02 eV) is due to
the electric fields which modify the core level binding energies and which were
not taken properly into account in Ref. \onlinecite{silvia2}.

As a last comment, we would like to come back to the structural
choices considered. In order to ascertain the dependence of the VBO
on the internal strain (not considered within MTE), we calculated the VBO
for superlattices having the experimental internal strain displacement:
we found a change in the band line-up of less than 0.02 eV (which
is well within our numerical accuracy).


\subsection{Results and discussion}
In Table \ref{tab2} we report our results for the interface contribution
$\Delta\:b$, the bulk contribution $\Delta\:E_b$ and the resulting band offset
value $\Delta\:E_v$. The superscripts $(nr)$ and $(r)$ indicate a
semirelativistic and fully relativistic
({\em i.e.} spin-orbit coupling included with a
perturbative approach) treatment of the valence
levels, respectively.  The final results for $\Delta\:E_v$$^{fin}$
include a correction for quasi-particle effects that were omitted in our self-
consistent calculations.  Thus, starting from the results of Ref.\
\onlinecite{louie}, we can include self-energy corrections
(considered uninfluenced by different strain conditions)
to the DFT-LDA energy levels  by adding the quantities
$\Delta^{QP}(GaSb/InSb)$ = -0.03 eV and $\Delta^{QP}(InAs/InSb)$ = +0.08 eV to
our calculated values .


In order to examine the effects of the ordering direction on these
quantities of
interest, we plot our Table I results in Figs. \ref{fig4} and \ref{fig5} for
GaSb/InSb and InAs/InSb interfaces, respectively, and also report the
calculated values for [001] systems with equal
pseudomorphic growth conditions. \cite{silvia2}
An inspection of Fig. \ref{fig4}
 (a) and Fig. \ref{fig5} (a) shows that,
as a function of the substrate lattice parameter,  the $\Delta\:b$
term is almost constant  in the GaSb/InSb [111] systems (as in all the [001]
structures \cite{silvia2} - see the dashed lines), whereas in the
InAs/InSb [111] systems the range in which
$\Delta\:b$ varies is slightly larger (about 0.06 eV).
This indicates that the charge
rearrangement at the interface is quite unaffected by strain conditions,
while it is much more influenced by the growth direction. In fact, we point
out that the $\Delta\:b$ absolute value is systematically higher (by as  much
as 0.15 eV) in the [111] structures than in those oriented along the [001]
axis;
furthermore, we observe that the trend relative to the small variations of the $
\Delta\:b$ term as a function of the substrate lattice parameter
seems to be opposite
in the two different orientations of the heterojunctions; in particular, as the
substrate lattice constant is increased, the $\Delta\:b$ increases (decreases)
in the [111] ([001]) case.
However, the variations of the $\Delta\:b$ term are of the same
magnitude as the error bar and it is therefore impossible to define a clear
$\Delta\:b$ trend as a function of strain conditions.

Let us now discuss the $\Delta\:E_b$ contribution to the band line-up.
As previously noticed in Ref. \onlinecite{silvia2}, the MTE predicts different
bond lengths (by as  much as 2 $\%$) perpendicular to the interface plane,
whether we consider the [001] or the [111] ordered
heterojunctions. This structural
difference is probably responsible for the different  $\Delta\:E_b$ term
in SLs grown on the same substrate (and therefore with the same in-plane strain
conditions)
but having different crystallographic orientation,
[001] or [111]  (see Fig. \ref{fig4} (b) and Fig. \ref{fig5} (b)).

Finally, we compare the valence band offsets for the CuAu and CuPt-like
heterojunctions.
The perfectly decreasing trend of the VBO
as $a_{sub}$ is increased, already evidenced for the [001] heterojunctions,
is still valid for those oriented along the [111] axis (see Fig. \ref{fig4}
 (c) and Fig. \ref{fig5} (c)).
Furthermore, we observe that the opportunity of tuning the band offset as a
function of the strain conditions is greater in [111] than in [001]
heterojunctions; in fact the range in which $\Delta\:E_v$ varies in going
from the smallest to the largest $a_s$ is about 0.8 eV for GaSb/InSb and
0.9 eV
InAs/InSb [111] interfaces.

\section{Conclusions}
We have presented  results obtained from FLAPW calculations performed for
[111] interfaces, in particular for GaSb/InSb and InAs/InSb heterojunctions.
In order to examine
the effects of the strain induced electric fields on the charge
redistribution (and hence on the potential profile)
near the heterojunction, we have proposed a simple scheme to locate the
interface region.  To a first approximation, we find that
piezoelectric effects lead to a planar distribution of
polarization charges, which doesn't affect the potential discontinuity.
We have followed an extrapolation procedure to obtain the
interface contribution to the VBO which
is found to be almost independent of the strain conditions, but is strongly
affected by the ordering direction.
The total potential discontinuity is found to vary by as much as 0.8 eV
for the GaSb/InSb and 0.9 eV for the InAs/InSb [111] interfaces, thus
confirming
strain  as an additional degree of freedom
to obtain ``{\em ad hoc}" band offsets.

\section{ACKNOWLEDGEMENTS}
We thank S. Massidda, R. Resta and B.W. Wessels
for stimulating discussions and/or a careful
reading of the manuscript.
Work at Northwestern University supported
by the MRL Program of the National Science Foundation, at the Materials
Research Center of Northwestern University, under Award No.
DMR-9120521,
and by a grant of computer time at the NSF
supported Pittsburgh Supercomputing Center.
Partial support by a supercomputing grant at Cineca (Bologna, Italy)
through the Consiglio Nazionale delle Ricerche (CNR) is also acknowledged.

%

\begin{table}
\centering
\small
\caption{Interface term ($\Delta\:b$), strained bulk term
($\Delta\:E_b$) and
valence band offset ($\Delta\:E_v$)  for (GaSb)$_3$/(InSb)$_3$
and (InAs)$_3$/(InSb)$_3$ [111] superlattices as a function of the
substrate lattice parameter neglecting
($\Delta\:E_b^{(nr)}$ and $\Delta\:E_v^{(nr)}$) and including
($\Delta\:E_b^{(r)}$ and $\Delta\:E_v^{(r)}$) spin-orbit
effects). $\Delta\:E_v^{(fin)}$ is the final result ($\Delta\:E_v^{(r)}+
\Delta^{QP}$), to be compared with experimental values (to the
best of our knowledge, not available
at present).
Energy differences
 are considered positive if the level relative
to the InSb layer is higher in energy with respect to
 the GaSb (InAs) layer
in the common-anion (common-cation) system.}
\vspace{5mm}
\begin{tabular}{|cl|r|cc|cc|c|}
\centering
& & $\Delta\:b$ & $\Delta\:E_b^{(nr)}$ & $\Delta\:E_b^{(r)}$ &
$\Delta\:E_v^{(nr)}$ & $\Delta\:E_v^{(r)}$ & $\Delta\:E_v^{(fin)}$ \\ \hline
&GaSb-subs.& +0.28 & +0.26 &+0.28&  +0.54 & +0.56 & +0.53\\
(GaSb)$_3$/(InSb)$_3$ &Av. subs. & +0.29 & -0.25 & -0.13 & +0.04 & +0.16&0.13 \\
&InSb-subs. & +0.30 & -0.70 & -0.54 & -0.40 & -0.24 & -0.27\\ \hline
&InAs-subs. & +0.16 & +0.85 & +0.99 & +1.01 & +1.15 & +1.23\\
(InAs)$_3$/(InSb)$_3$ &Av. subs. & +0.19 & +0.30 & +0.51 & +0.49 & +0.70&
+0.78\\
& InSb-subs. & +0.22 & -0.20 & +0.01 & +0.02 & +0.23& +0.31\\
\end{tabular}
\label{tab2}
\end{table}

\begin{figure}
\caption{Structure for the [111] GaSb/InSb 3x3 superlattice.
The atoms
denoted by ``b$_1$", ``b$_2$" and ``i" indicate the atoms in the two bulk
and interface regions, respectively. The
dotted lines $\pi_c$ and $\pi_s$ denote the conventional positions of the
interface planes at the center and at the sides of the unit cell, respectively.}
\label{fig1}
\end{figure}

\begin{figure}
\caption{Sb 1s core levels binding energies ($E^b_{CL}$) (empty circles)
in the (a) 3x3 and (b) 4x4 GaSb/InSb systems (grown on an average
substrate) as a function of the $z$ coordinate
perpendicular to the interface.
The energy scale is referred to an arbitrary zero.
The dashed lines indicate the linear
interpolation of the bulk core levels energies;
their intercepts with the conventional interface
planes (full diamonds) define the interface contribution
($\Delta\:b$) to the VBO.}
\label{fig2}
\end{figure}

\begin{figure}
\caption{In 1s core levels binding energies ($E^b_{CL}$)
in the (a) 3x3 and (b) 4x4 InAs/InSb systems (grown on an average
substrate) as a function of the $z$ coordinate
for [111] (empty circles)
and [001] (empty squares) ordered SLs.
Labels and energy scale as in Fig. 2.}
\label{fig3}
\end{figure}

\begin{figure}
\caption{Interface contribution $\Delta\:b$ (panel (a)), bulk  contribution
($\Delta\:E_b^{(r)}$) (panel (b))
and total valence band offset ($\Delta\:E_v^{(r)}$) (panel (c))
for GaSb/InSb interfaces
as a function of the substrate lattice constant.
Filled (empty) symbols  and solid (dashed) lines refer to
[111] ([001]) ordered heterojunctions.}
\label{fig4}
\end{figure}

\begin{figure}
\caption{Interface contribution ($\Delta\:b$) (panel (a)), bulk  contribution
($\Delta\:E_b^{(r)}$) (panel (b))
and total valence band offset ($\Delta\:E_v^{(r)}$) (panel (c))
for InAs/InSb interfaces
as a function of the substrate lattice constant.
Filled (empty) symbols  and solid (dashed) lines  refer to
[111] ([001]) ordered heterojunctions.}
\label{fig5}
\end{figure}


\begin{thebibliography}{100}

\bibitem{maria} N.Tit, M.Peressi and S.Baroni, Phys. Rev. B {\bf 48}, 17607
(1993).

\bibitem{gorczyca} I. Gorczyca and N. E. Christensen, Phys. Rev. B
{\bf 48}, 17202 (1993).

\bibitem{swaz} Su-Huai Wei and Alex Zunger, Phys. Rev. B {\bf 52},
12039 (1995).

\bibitem{ohler} C.Ohler {\em et al.}, Phys. Rev. B {\bf 50}, 7833 (1994).

\bibitem{silvia2} S.Picozzi, A.Continenza and A.J.Freeman, Phys. Rev. B
{\bf 53}, 10852 (1996).

\bibitem{vandewalle} C. G. Van de Walle and R. M. Martin, Phys. Rev. B {\bf 35},
8154 (1987).

\bibitem{FLAPW}
H.J.F.Jansen and A.J.Freeman, Phys. Rev. B {\bf 30}, 561 (1984);
M.Weinert, H.Krakauer, E.Wimmer and A.J.Freeman,
{\em ibid.} {\bf 24}, 864 (1981).

\bibitem{dft} P.Hohenberg and W. Kohn, Phys. Rev.
 {\bf 136}, B864 (1984); W.Kohn and L.J.Sham, {\em ibid.}{\bf 145}, 561
 (1966).

\bibitem{froyen} S. Froyen, A. Zunger and A. Mascarenhas, Appl. Phys. Lett.
{\bf 68}, 2852 (1996)

\bibitem{harshman} P. J. Harshman and S. Wang, J. Appl. Phys. {\bf 71}, 5531
(1990).

\bibitem{bbr}S. Baroni, R. Resta and A. Baldereschi, Phys. Rev. Lett. {\bf 61},
734 (1988).

\bibitem{martin} R. M. Martin, Phys. Rev. B {\bf 5}, 1607 (1972).

\bibitem{degiro} S. De Gironcoli, S. Baroni and R. Resta, Phys. Rev. Lett. {\bf
 62}, 2853 (1989).

\bibitem{nye} J. F. Nye, ``{\em Physical Properties of Crystals}", (Oxford
University, New York, 1985).

\bibitem{dalcorso} A. Dal Corso, R. Resta and S. Baroni, Phys. Rev. B {\bf 47},
16252 (1993).

\bibitem{mailhotrmp} D. L. Smith and C. Mailhot, Rev. Mod. Phys. {\bf 62},
173 (1990).

\bibitem{silvia1} S.Picozzi, A.Continenza and A.J.Freeman, Phys. Rev. B
{\bf 52}, 5247 (1995).

\bibitem{martin1} R. M. Martin, Phys. Rev. B {\bf 1}, 4005 (1970).

\bibitem{vdw2} C. Van de Walle and R. M. Martin, Phys. Rev. B
{\bf 34}, 5621 (1986).

\bibitem{nota} From FLAPW total energy and atomic force calculations of
full atomic relaxations in some of the strained binary systems,
we found relaxed bond lengths differing
among themselves at most by 0.6 $\%$
and with only a 0.4 $\%$ disagreement with the MTE predictions. This is
probably due to the great sensitivity of the internal strain
parameter $\zeta$ on the lattice constant (see for example
Refs. \onlinecite{degiro,martin1}).

\bibitem{HL}  L.Hedin and B.I.Lundqvist, J. Phys. C. {\bf 4}, 2064 (1971).

\bibitem{ale90} A.Continenza, S.Massidda and A.J.Freeman, Phys. Rev. B
{\bf 42}, 3469 (1990).

\bibitem{harrison} W. A. Harrison , ``{\em Electronic Structure and the
Properties
of Solids"} (W. H. Freeman \& Co., San Francisco, 1980).


\bibitem{vboaz} Su-Huai Wei and Alex Zunger,
J. Vac. Sci. Technol. {\bf B 5 (4)}, 1239 (1987).

\bibitem{mmf} S. Massidda, B. I. Min and A. J. Freeman,
Phys. Rev. B {\bf 35}, 9871 (1987) and references therein.

\bibitem{fiore} A. Satta, V. Fiorentini, A. Bosin, F. Meloni and D.
Vanderbilt,
to be published.

\bibitem{louie} X. Zhu and S. G. Louie, Phys. Rev. {\bf 43}, 14142 (1991).


\end{thebibliography}
\end{document}